\documentclass[12pt]{iopart}
\usepackage{iopams}
\usepackage{epsfig}

\begin{document}
%----------------------------------------------------------------------------
%                              TITLE
%----------------------------------------------------------------------------
\title[Interplay of topological and structural defects in the $2D$ $XY$ model]{
Interplay of topological and structural defects in the $2D$ $XY$ model
}
%----------------------------------------------------------------------------
%                              AUTHORS
%----------------------------------------------------------------------------

\author{O.\ Kapikranian$^{1,2}$, B.\ Berche$^{2}$, and
 Yu.\ Holovatch$^{1,3}$}

\address{$^{1}$
Institute for Condensed Matter Physics, National Acad. Sci. of Ukraine, UA-79011 Lviv, Ukraine}

\address{$^{2}$ Laboratoire de Physique des Mat\'eriaux, UMR CNRS 7556,
  Universit\'e Henri Poincar\'e, Nancy 1, F-54506  Vand\oe uvre les Nancy Cedex, France}

\address{$^{3}$
Institut f\"ur Theoretische Physik, Johannes Kepler Universit\"at
Linz, A-4040 Linz, Austria}

%----------------------------------------------------------------------------
%                             ABSTRACT
%----------------------------------------------------------------------------
\begin{abstract}

The present work is devoted to the investigation of the
interaction between vortices (topological defects) and
site-impurities (structural defects) in the $2D$ $XY$ model and
its influence on the well-known properties of the pure system. The
main goal is a theoretical description of the
Berezinskii-Kosterlitz-Thouless (BKT) temperature reduction by
quenched non-magnetic impurities, based on the vacancy-vortex
interactions and the vortex-pair dissociation mechanism of the
transition. The non-magnetic impurity interaction with a system of
vortices can be found either from the phenomenological theory of topological
defects or from the Villain model. We take both paths
and compare the results obtained. Our prediction for the BKT
temperature reduction is confirmed by the available Monte Carlo
data.
\end{abstract}
%\submitto{\JPA}

\eads{\mailto{kapikran@lpm.u-nancy.fr},
\mailto{berche@lpm.u-nancy.fr},
\mailto{hol@icmp.lviv.ua}}

\maketitle

\section{Introduction}\label{Intro}

An object of our interest is the two-dimensional $XY$ model, described
by the Hamiltonian:

\begin{equation}\label{XY_Hamilt}
H = -J\sum_{\left<{\bf r,r'}\right>}\left(S^x_{\bf r}S^x_{\bf
r'}+S^y_{\bf r}S^y_{\bf r'}\right)\ .
\end{equation}
Here, ${\bf S}_{\bf r}$ are unit spins placed on the sites of the square 
lattice with the nearest neighbour interaction $J$. 
This model is well known for its extraordinary properties connected with the presence of
topological defects (vortices). It can also be considered as the limiting case of the
$2D$ easy-plain Heisenberg model: $H_{\mathrm{e.p.}} = -J\sum_{<{\bf r,r'}>}
\left(S^x_{\bf r}S^x_{\bf r'}+S^y_{\bf r}S^y_{\bf r'}+\lambda 
S^z_{\bf r}S^z_{\bf r'}\right)$, when $\lambda = 0$.
It has been argued that quasi-two-dimensional types of real 
magnetic materials, like layered magnets or ultrathin 
magnetic films 
can be satisfactory described within the $2D$ easy-plane Heisenberg
model~\cite{HikamiTsuneto80}, and since
the behaviour of this model has been found qualitatively
similar to that of the $2D$ $XY$ model in a rather wide range
of anisotropy parameter~\cite{Gouvea89,KawabataBishop82}, it seems natural to
use the $2D$ $XY$ model as a suitable device for the study of real quasi-$2D$ 
magnets. Thus the question of the influence of impurities, always present in
the lattice structure of real materials, should be and has been
posed in recent years~\cite{PereiraEtAl03,Berche03,WysinEtAl05}.

We define here the $2D$ $XY$ model without the non-interacting (and thus
unimportant) component $S^{z}_{\bf r}$ in the trace of the
system, as is often done in literature. This case is also referred to by some
authors as the planar rotator model (see, for example,~\cite{Mol06}). This should not confuse
the reader, since the non-interacting component would not change the 
qualitative picture anyway.

The Hamiltonian (\ref{XY_Hamilt}) can be written in a more
convenient form for calculation in terms of the angle variables
$-\pi<\theta_{\bf r}\le \pi$:

\begin{equation}\label{Ham_XY}
H =  - J\sum_{\bf r}\sum_{\alpha=x,y}\cos(\theta_{{\bf
r+a}_\alpha} - \theta_{\bf r})
\end{equation}
where ${\bf a}_x = (a,0)$ and ${\bf a}_y = (0,a)$ form an elementary
basis of the lattice, and $J$ is the ferromagnetic coupling. Then,
since we want to study a system with non-magnetic impurities
(vacancies) in the lattice we introduce the ``occupation numbers":

\begin{equation}
c_{\bf r} = \left\{ \begin{array}{ll}
1, & \textrm{if there is a spin;}\\
0, & \textrm{if the site is empty.}
\end{array} \right.\label{OccNumb}
\end{equation}
and construct the Hamiltonian:

\begin{equation}\label{Ham_XYdil}
H =  - J\sum_{\bf r}\sum_{\alpha=x,y}\cos(\theta_{{\bf
r+a}_\alpha} - \theta_{\bf r})c_{\bf r}c_{\bf r'}\ .
\end{equation}

The introduction of such impurities (in a general case) makes the
model impossible to diagonalize in the spin-wave approximation by
a Fourier transformation as it is possible to do for the regular
(without structural defects) lattice~\cite{Wegner67}. One
distinguishes quenched and annealed types of dilution. Annealed
dilution is understood as impurities being in thermodynamical
equilibrium with the spin degrees of freedom, so the averaging
over the occupation numbers (\ref{OccNumb}) should be taken
already in the partition function~\cite{Brout59}. In the quenched
dilution case, the impurities are frozen at their positions with
some fixed probability and one should average
the observable quantities 
(like the spin-spin
correlation function or the free energy of the system)
over different
configurations, and  not the
partition function itself. The latter statement, formulated
in~\cite{Brout59}, was subsequently rigorously proven
in~\cite{Mazo63}.

The random (quenched) dilution means that probability to remove a spin from a
site is fixed and independent on the other sites state. So the averaging for all the possible configurations of vacancies can be written as

\begin{equation}\label{ConfAvrg}
\overline{(...)} = \sum_{\{c_{\bf r}=0,1\}}\ P(\{c_{\bf
r}\})(...)\ ,
\end{equation}
with the probability function

\begin{equation}
P(\{c_{\bf r}\})\ =\ \prod_{\bf r}[c\delta_{c_{\bf r},1}+(1-c)\delta_{c_{\bf r},0}]\ .
\end{equation}
This distribution is set in such a way that we obtain in average a
system with  concentration of magnetic sites $c$ (or  fraction of
impurities $1-c$).

Although the case of annealed impurities seems to be well studied
and clear enough~\cite{BerkerNelson79,CardiScalapino79,Gruber02},
the influence of quenched dilution is a problem for which there
are still unsolved questions and which deserves attention. For
example, up to our knowledge, so far there are only Monte Carlo
results for the phase diagram
$(T_{\mathrm{BKT}},c)$~\cite{Berche03,WysinEtAl05}, showing the
reduction of $T_{\mathrm{BKT}}$ with decreasing concentration $c$
of the magnetic component, and no theoretical constructions trying
to explain this reduction. Also an approach to
investigate the diluted model in the spin-wave approximation has
been proposed in~\cite{Kapikranian07}.

It is well known that the Berezinskii-Kosterlitz-Thouless
transition is driven by the topological defects and is in some
sense equivalent to the neutral $2d$ Coulomb gas transition to
conducting state~\cite{Minnhagen87,KosterlitzThoulessBerezinskii}.
In the present paper we describe the critical
temperature reduction by an analysis based
on the vacancy-vortex interactions (Section \ref{II}).
The form of this interaction can be found either from the phenomenological theory
of topological defects~\cite{Nelson02,Chaikin95} or directly from
the Villain model~\cite{Villain75} which can be regarded as a low
temperature approximation of the $2d$ $XY$ model~\cite{JoseEtAl} 
(Sections \ref{I} and \ref{III} respectively).

\section{The vacancy interaction in a system with vortices}\label{I}

An efficient way to study vortices in the $2D$ $XY$ model is a
continuous elastic medium approximation
where the spin-wave excitations are
forgotten and the topological defects are obtained from the
``elastic" energy minimization under some special topological
constraints. The spin variables $\theta_{\bf r}$ defined on the
sites of the initial lattice are promoted to a continuous field
$\theta({\bf r})$, and the continuous limit of the spin-wave
(harmonic) approximation of the Hamiltonian (\ref{Ham_XY}) is
taken as the ``elastic" energy of the system:

\begin{equation}\label{Enrg_Elast}
E_{\mathrm{el}}\ =\ \frac 1 2 J\sum_{\bf r}\sum_{\alpha=x,y}
(\Delta_\alpha\theta_{\bf r})^2\ \simeq\ \frac 1 2 J \int d{\bf
r} (\nabla\theta({\bf r}))^2\ .
\end{equation}
The configuration of the field $\theta({\bf r})$ that satisfies
the topological condition $\oint d \theta\ =\ 2\pi q$
(definition of a vortex with winding number $q$), 
where the integral 
is over an arbitrary path enclosing the point defined
as the vortex center, and has the
minimal elastic energy (\ref{Enrg_Elast}), can be written in a polar coordinates system
(centered at the center of the vortex) as:

\begin{equation}\label{vort_eq}
\theta = q\varphi+\mathrm{const}\ ,
\end{equation}
and its gradient,
${\displaystyle{\nabla\theta=\frac{q}{r}(-\sin\varphi,\cos\varphi)}}$,
can be found easily. This gradient is always perpendicular to the
radius-vector of the point drawn from the origin. The configuration
obtained
is called a vortex with the 
charge (winding number) $q$. The vortex is completely set by its charge,
the constant in (\ref{vort_eq}) is absolutely arbitrary, since one can switch 
from a configuration with one constant to a configuration with another 
constant without changing the energy (although the field 
configuration visually depends on the value of the constant).

Actually, the total energy of such a configuration 
can not be correctly expressed by (\ref{Enrg_Elast}), since in the 
continuous
limit we have a singularity in the center of the vortex. Due to this, one 
has to specify
the core energy of the vortex which is always finite and the elastic energy
becomes:

\begin{displaymath}
E_{\mathrm{el}}^{\mathrm{pure}}\ =\ q^2 J\pi \int_{A}^{L}
\frac{dr}{r}\ =\ q^2 J\pi\ln(L/A)\ .
\end{displaymath}
It is divergent with the system size $L$ and  $A$ is the
radius of the core. We do not touch here the nontrivial question about the size of
this core region and its energy estimation~\cite{Chaikin95}.

The elastic energy of a vortex with a non-magnetic vacancy at some
sufficient distance $r$ from the center can be found as the 
energy that corresponds to the four
bonds removed (square lattice) subtracted from the energy of the pure system:

\begin{eqnarray}
E_{\mathrm{el}}^{\mathrm{dil}} &=& E_{\mathrm{el}}^{\mathrm{pure}} - E_{\mathrm{vac}} 
= E_{\mathrm{el}}^{\mathrm{pure}} - \frac{1}{2}J\sum_{\alpha=x,y}
[(\nabla\theta\cdot{\bf a}_{\alpha})^2+(-\nabla\theta\cdot{\bf a}_{\alpha})^2]
\\\nonumber
&=& E_{\mathrm{el}}^{\mathrm{pure}} - \frac{1}{2} q^2 J\frac{a^2}{r^2}
\left\{2\sin^2\varphi+2\cos^2\varphi\right\}\
=E_{\mathrm{el}}^{\mathrm{pure}} - Jq^2\left(a/r\right)^2\ .
\end{eqnarray}
Thus, a non-magnetic vacancy has an
attractive interaction with either a 
positive or a negative vortex charge.
This is in good agreement
with references~\cite{PereiraEtAl03,Wysin03}.
Of course, this result is obtained via the assumption
that the
vacancy does not disturb the vortex configuration, 
an hypothesis which was reliably
argued in~\cite{PereiraEtAl03}. Our result is almost equivalent to
that found in the paper mentioned, but seems to be a bit more
definite since in~\cite{PereiraEtAl03} the coefficient in the
interaction depends on the way of cutting out an area of the
continuous field around the vacancy, and in our case it is only a
matter of the lattice structure.

We go further and consider a vortex pair (winding numbers $q$,
$q'$) containing an impurity. The spin configuration of the pair
is simply the superposition of the single-vortex fields:
$\theta({\bf r})+\theta'({\bf r})$ and thus the gradient is
$\nabla\theta+\nabla\theta'$. For the pure system the simple 
integration gives the elastic energy of such a pair:

\begin{equation}
E_{\mathrm{el}}^{\mathrm{pure}} = -2\pi Jqq'\ln(R/A) + \pi J(q+q')^2\ln(L/A)\ ,
\end{equation}
where $R$ is the distance between the two vortices. Note that 
the second term, which is divergent, vanishes for a 
neutral pair ($q'=-q$).

Let the polar coordinates of the
impurity be $(r,\varphi)$ in the coordinate system centered on
the vortex $q$ and $(r',\varphi')$ in the system centered on the
second vortex $q'$. We write down the result for the energy associated
with this vacancy:

\begin{equation}\label{Vac_and_pair}
E_{\mathrm{vac}}
= Ja^2\left((q/r)^2+ (q'/r')^2 +
2(q/r)(q'/r')\cos(\varphi-\varphi')\right)\ .
\end{equation}

For a system with an
arbitrary number of vortices and a vacancy at the point ${\bf r}$
we can generalize the elastic energy as:

\begin{eqnarray}\label{ManyVort}
E_{\mathrm{el}}^{\mathrm{dil}} &=& E_{\mathrm{el}}^{\mathrm{pure}} - E_{\mathrm{vac}}({\bf r}) 
\\\nonumber
&=& E_{\mathrm{el}}^{\mathrm{pure}} - J\sum_{\bf R}\sum_{\bf R'} q({\bf
R})q({\bf R'})\frac{a^2}{|{\bf R-r}||{\bf R'-r}|}\frac{({\bf
R-r})({\bf R'-r})}{|{\bf R-r}||{\bf R'-r}|}\ ,
\end{eqnarray}
where the sums span all the topological defects present in the
system.

So far we have been considering only one impurity in the system.
Of course, a single vacancy does not have any influence on an
infinite system, but, having formula (\ref{Vac_and_pair}) for the
vortex-pair-vacancy interaction, we can pass to the case of a
finite fraction of empty sites and make some conclusions about the
critical temperature behaviour with the concentration $c$. This
will be the subject of the following section.

\section{The critical temperature reduction by quenched dilution}\label{II}

In the previous section we obtained the form of the spinless site
interaction with a system of topological defects. Instead
of this, one considers now a single vortex 
pair in a system with some
fraction of spins removed. The elastic energy can be written as
\begin{equation}\label{Imp_Sum}
E_{\mathrm{el}}^{\mathrm{dil}} = E_{\mathrm{el}}^{\mathrm{pure}} - \sum_{\bf
r_\mathrm{{vac}}}E_{\mathrm{vac}}({\bf r})\ ,
\end{equation}
where $E_{\mathrm{el}}^{\mathrm{pure}}$ is the pure system energy,
$E_{\mathrm{vac}}({\bf r})$ is the energy associated with a
vacancy at the point ${\bf r}$, given by
(\ref{Vac_and_pair}), and the sum is over all the vacancies in the
system.

Obviously (\ref{Imp_Sum}) is not exact because there
are certainly some impurities which occupy neighbouring sites and
thus destroy smaller numbers of 
links per vacancy. Their contribution
to the energy will be different, but (\ref{Imp_Sum}) can be
considered as a reasonable approximation when the concentration of
dilution is weak enough. Eq.(\ref{Imp_Sum}) can be equally written
in the continuous approximation:
\begin{equation}\label{Enrg_Dil}
E_{\mathrm{el}}^{\mathrm{dil}} = E_{\mathrm{el}}^{\mathrm{pure}} - \int d{\bf
r}\rho_{\mathrm{vac}}({\bf r})E_{\mathrm{vac}}({\bf r})\ ,
\end{equation}
with the impurity density introduced as
\begin{equation}\label{imp_density}
\rho_{\mathrm{vac}}({\bf r}) = \sum_{\bf r'}\delta({\bf
r-r}')(1-c_{\bf r})\ ,
\end{equation}
here $\delta$ is for a delta-function and $c_{\bf r}$-s are the
occupation numbers (\ref{OccNumb}).

The energy (\ref{Enrg_Dil}) can serve to estimate the transition
temperature, $T_{\mathrm{BKT}}$. Consider an ideal system that is
constituted of a single neutral pair of vortices with winding
numbers of modulus $1$, thus one has only one degree of freedom, the
separation $R$ between the vortices
One can define $T_{\mathrm{BKT}}$ as the
temperature when this pair dissociates~\cite{Izyumov88}, 
i.e. the thermodynamical
average

\begin{equation}\label{R_square}
\left<R^2\right> = \frac{\int_{a}^{\infty}R^3 e^{-\beta
E_{\mathrm{el}}(R)}dR}{\int_{a}^{\infty}R e^{-\beta
E_{\mathrm{el}}(R)}dR}
\end{equation}
goes to infinity. With the undiluted system it happens at
$kT_{\mathrm{BKT}}/J \simeq \pi/2$, since
$E^{\mathrm{pure}}_{\mathrm{el}}(R) = 2\pi J\ln(R/a)$ and one
easily finds $\left<R^2\right> = a^2(\pi\beta J - 1)/(\pi\beta J -
2)$. Using the best present estimate of the BKT
temperature, $kT_{\mathrm{BKT}}/J \simeq
0.893$~\cite{HasenbuschPinn97}, 
it is obvious that this
calculation gives quite a rough result.
Nevertheless, in spite of a quantitative vagueness, this
approach is qualitatively correctly based on the BKT transition mechanism.
We choose to use it due to its
simplicity and expect it to be satisfactory to examine the
influence of non-magnetic dilution on the critical temperature.

Now, with the energy (\ref{Enrg_Dil}) of a vortex-antivortex
system with spin dilution, we can search the BKT point as the
temperature where (\ref{R_square}) diverges. With
$\rho_{\mathrm{vac}}$ defined for an arbitrary configuration of
impurities as (\ref{imp_density}) it is quite complicated, however
one can use some approximate form of the density, reflecting its
essential features. Here the quenched and annealed dilutions
should be discriminated. In the frozen case the spins are removed
randomly with the same probability for each site, so there is no
preference for any part of the lattice to be more or less diluted
than the rest of the system. 
Of course, fluctuations of random nature rather than 
thermal origin exist. The
probability for these fluctuations goes 
to zero as the size of the lattice increases to
infinity. Based on these arguments,
a perturbation expansion has been
proposed~\cite{Kapikranian07} where the 0th order can be
considered as a ``perfectly homogeneous" dilution. Here we get the
corresponding approximation replacing (\ref{imp_density}) with a
``smeared" density:

\begin{displaymath}
\rho({\bf r}) \simeq
(1-c)N/(a^2N) = (1-c)/a^2\ ,
\end{displaymath}
which is simply the number of empty sites divided by the total
volume. Now the integral in (\ref{Enrg_Dil}) can be simply
calculated and gives $E_{\mathrm{el}}(R) = [1-2(1-c)] 2\pi
J\ln(R/a)$. It follows that the BKT temperature is just
$kT^{\mathrm{dil}}_{\mathrm{BKT}}/J = [1-2(1-c)]\pi/2$, or,
normalizing to the pure model critical temperature,

\begin{equation}\label{TdilByTpure}
T^{\mathrm{dil}}_{\mathrm{BKT}}/T^{\mathrm{pure}}_{\mathrm{BKT}}=
1-2(1-c)\ .
\end{equation}

The critical temperature decreases with the dilution concentration, as one naturally
expects due to the decrease of the average coordination number. Moreover, 
although our derivation was based on the assumption of
a weak dilution, formula (17) predicts a vanishing of the $T_{\mathrm{BKT}}$ at
$c=0.5$. Being qualitatively correct, the last estimate differs from
the known site percolation threshold concentration on a square lattice
$c\simeq 0.59$ \cite{Gebele84}.

\begin{figure} [h]
  \centerline{\includegraphics[width=13.5cm]{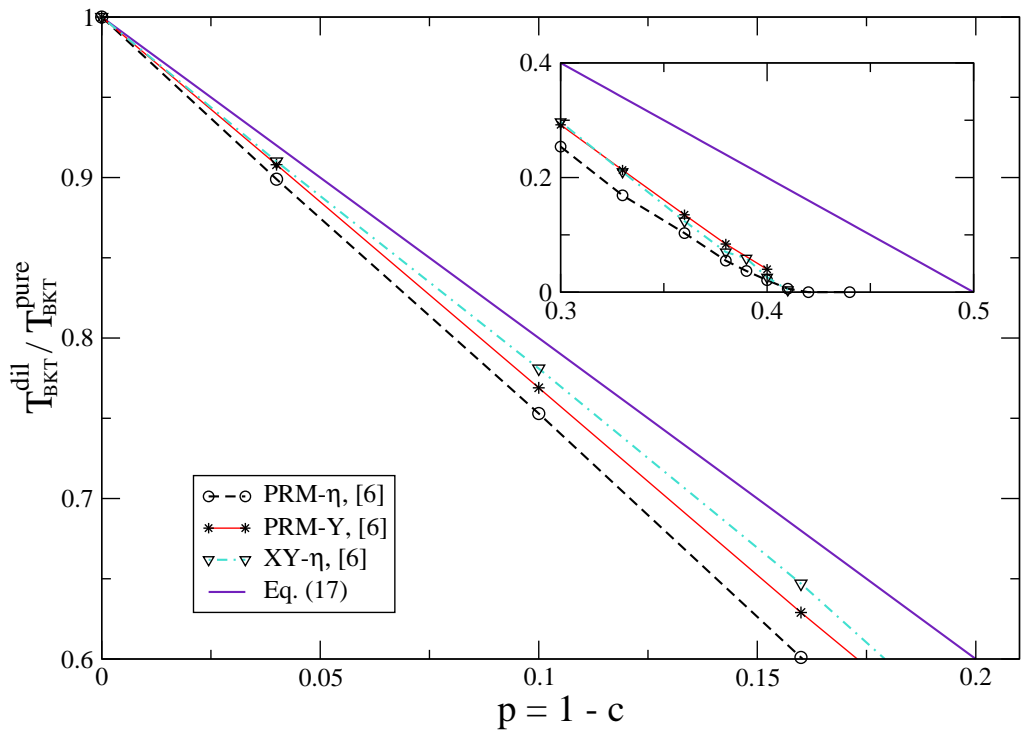}}
        \caption{The phase diagram of the $2D$ $XY$ model with the quenched dilution of concentration $p=1-c$ ($c$ is the concentration of magnetic sites). The Monte Carlo simulation results of~\cite{WysinEtAl05} are compared 
        with our theoretical prediction (\ref{TdilByTpure}). The insert shows the vicinity of the percolation threshold.}
        \label{Fig1}  \vskip -0cm
\end{figure}

While the influence of quenched dilution in particular spin models,
for example 
in the $2D$ and $3D$ Ising model~\cite{Shalaev94,Folk03}, 
is well studied in numerous Monte Carlo
simulations, for the model under consideration the computer experiment
results are rather poor.
We compare our result (\ref{TdilByTpure}) with the available Monte Carlo
data for the diluted two-dimensional $XY$ model~\cite{WysinEtAl05} (Fig.1).
The simulations were performed for the $XY$ and planar rotator models 
with quenched dilution in two dimensions. Note that in terms of the paper~\cite{WysinEtAl05} 
our model (\ref{Ham_XYdil}) is the planar rotator model (PRM). Results
of~\cite{WysinEtAl05} for the PRM critical temperature estimated from two 
different methods, by the helicity modulus jump (PRM-Y) and by the spin
correlation function exponent $\eta$ behaviour (PRM-$\eta$), differ
essentially in the region of weak dilution. The $XY$ model points are
again different. Although all three MC results go eventually to the 
correct percolation threshold (see the insert in Fig.1) there are few low 
concentration points and they do not
seem to be reliable enough to make some
strong conclusion about our result. However, at least the linear 
character of equation~(\ref{TdilByTpure})
seems to be present in all  three MC sets 
in the weak dilution range, and this observation is also
supported by~\cite{Berche03}.

\section{The Villain model with nonmagnetic impurities}\label{III}

According to~\cite{JoseEtAl} the Villain model can be derived
in the low temperature limit from the Hamiltonian

\begin{equation}\label{HamVillDeriv}
H = - J \sum_{\left<{\bf r, r'}\right>}\left[\ \cos(\theta_{\bf
r}-\theta_{\bf r'})-1\ \right]
\end{equation}
which is equivalent to that of the $2D$ $XY$ model (\ref{Ham_XY}).
We apply here the scheme of this derivation to the case of a diluted model,
starting with the Hamiltonian

\begin{equation}\label{HamVillDerivDil}
H = - J \sum_{\left<{\bf r, r'}\right>}\left[\ \cos(\theta_{\bf
r}-\theta_{\bf r'})-1\ \right]\ c_{\bf r}c_{\bf r'}\ ,
\end{equation}
with $c_{\bf r}$-s being the occupation numbers (\ref{OccNumb}). The partition
function of the model is then written:

\begin{equation}\label{PartFunc-1}
Z = \left(\prod_{\bf r}\int_{-\pi}^{\pi} \frac{d\theta_{\bf
r}}{2\pi}\right)\ \exp\left[\sum_{\left<{\bf r,
r'}\right>}V(\theta_{\bf r}-\theta_{\bf r'})\ c_{\bf r}c_{\bf
r'}\right]\ ,
\end{equation}
with $V(\theta) = K \left[\ \cos\theta-1\ \right]$ and $K=J/k_BT$.

The goal of the derivation is to
find an approximate form of the expression under the integral in
(\ref{PartFunc-1}), which preserves its initial symmetry and is easier to
integrate. Namely it will be a 
superposition of exponents with quadratic arguments like in the SWA, 
but some new descrete variables will appear which subsequently can be 
associated with the vortex excitations in the system. 
Going through this procedure with the impurity variables $c_{\bf r}$-s 
one can expect to find the influence of  dilution on 
the vortex energy contribution.

As the first step one has to decompose the Boltzmann 
factor in (\ref{PartFunc-1}) in Fourier series:

\begin{equation}\label{FourierExp}
\exp\left[\sum_{\left<{\bf r,
r'}\right>}V(\theta_{\bf r}-\theta_{\bf r'})\ c_{\bf r}c_{\bf
r'}\right] =
\prod_{\left<{\bf r,r'}\right>}\sum_{s=-\infty}^{+\infty} \Theta(s)\
e^{is(\theta_{\bf r}-\theta_{\bf r'})}\
e^{\widetilde{V}(s)c_{\bf r}c_{\bf r'}}\ ,
\end{equation}
where $\Theta(s) =  c_{\bf r}c_{\bf r'} + (1-c_{\bf r}c_{\bf
r'})\delta_{s,0}$ naturally appears to ensure the equality when $c_{\bf r}c_{\bf r'}=0$.
The Fourier variable $s$ depends on  two real space variables: $s=s({\bf r,r'})$.

Now, applying the Poisson summation formula~\cite{MorseFeshbach53}:

\begin{equation}\label{Poisson}
\sum_{s=-\infty}^{\infty}g(s) =
\sum_{m=-\infty}^{\infty}\int_{-\infty}^{\infty}d\phi\ g(\phi)\
e^{-2\pi i\phi m}\ ,
\end{equation}
usually used to improve the convergence of the series, to each of the sums in (\ref{FourierExp}), one can rewrite the partition function as:

\begin{equation}\label{PartFunc-2}
Z = \left(\prod_{\bf r}\int_{-\pi}^{\pi} \frac{d\theta_{\bf
r}}{2\pi}\right)\sum_{m_{\bf r,r'}=-\infty}^{+\infty}\Theta(m_{\bf r,r'})\
e^{\sum_{\left<{\bf r, r'}\right>}V_0(\theta_{\bf r}-\theta_{\bf
r'}-2\pi m_{\bf r,r'})\ c_{\bf r}c_{\bf r'}}\ ,
\end{equation}
with $e^{V_0(\theta)} = \int_{-\infty}^{\infty} d\phi\
e^{\widetilde{V}(\phi)}\ e^{i\phi\theta}$. So far, no special assumption 
have been made and the result above is exact.

Now let us consider the low temperature limit.
In this approximation it is not difficult to find that $e^{V_0(\theta)}
\simeq e^{-K\theta^2/2}$. The latter comes as the result of the
asymptotic form ($K\to\infty$): $e^{\widetilde{V}(s)} =
\frac{1}{2\pi}\int_{0}^{2\pi} d\theta\ e^{-is\theta}\
e^{K(\cos\theta-1)} \approx \frac{1}{\sqrt{2\pi K}}\
e^{-s^2/(2K)}$. One obtains the partition function

\begin{equation}\label{PartFunc-3}
Z = \sum_{m_{\bf r,r'}=-\infty}^{+\infty}\Theta(m_{\bf r,r'})\left(\prod_{\bf
r}\int_{-\pi}^{\pi} \frac{d\theta_{\bf r}}{2\pi}\right)\
e^{-K\sum_{\left<{\bf r, r'}\right>}(\theta_{\bf r}-\theta_{\bf
r'}-2\pi m_{\bf r,r'})^2 c_{\bf r}c_{\bf r'}}
\end{equation}
of the desirable form, but as far as the limits of integration
remain $(-\pi, \pi)$ all terms with $m\neq 0$ give vanishing
contribution (as $K\to\infty$) and (\ref{PartFunc-3}) is equivalent to the SWA. To
repair this, extending the limits of integration, one can use the equality:

\begin{equation}\label{INFTY}
\int_{-\pi}^{\pi}d\varphi f(\varphi) = \lim_{\varepsilon\to
0}2\sqrt{\beta\pi\varepsilon} \int_{-\infty}^{\infty}d\varphi\
e^{-\beta\varepsilon\varphi^2}f(\varphi)\ ,
\end{equation}
true for any periodic function $f(\varphi) = f(\varphi+2\pi)$. Eq.(\ref{INFTY})
can be easily checked by passing to the Fourier transform:
$f(\varphi) = \sum_{s=-\infty}^{\infty}\ e^{is\varphi}F(s)$. The
left part of (\ref{INFTY}) is just $2\pi F(0)$. Integrating term
by term  the right part of (\ref{INFTY}) and taking the limit
$\varepsilon\to 0$ one finds again $2\pi F(0)$.

Finally, one has the partition function that describes the Villain
model with non-magnetic impurities:

\begin{equation}\label{PartFunc-4}
Z = \sum_{m({\bf r,r'})=-\infty}^{+\infty}\Theta(m_{\bf r,r'})\left(\prod_{\bf
r}\int_{-\infty}^{\infty} \frac{d\theta_{\bf r}}{2\pi}\right)\
e^{-\beta H_{\mathrm{Vill}}^{\mathrm{dil}}},
\end{equation}
with the Hamiltonian

\begin{equation}\label{Hamilt_Villain}
H_{\mathrm{Vill}}^{\mathrm{dil}} = J \sum_{\left<{\bf r,
r'}\right>}(\theta_{\bf r}-\theta_{\bf r'}-2\pi m_{\bf r,r'})^2
c_{\bf r}c_{\bf r'} + \varepsilon\sum_{\bf r}\theta_{\bf r}^2 \ .
\end{equation}
When all the $c_{\bf r}$-s are taken equal $1$ it turns to the Hamiltonian of the regular Villain
model~\cite{Villain75}.

It is known that the pure Villain model Hamiltonian can be divided
into two parts: one corresponding to the spin-wave excitations and
another one that corresponds to the vortex contribution. It is achieved by the
Fourier transformation of the spin variables,

\begin{displaymath}
\theta_{\bf r} = \frac{1}{\sqrt{N}}\sum_{\bf k}\ e^{-i{\bf
kr}}\theta_{\bf k}\ ,\qquad \theta_{\bf k} =
\frac{1}{\sqrt{N}}\sum_{\bf r}\ e^{i{\bf kr}}\theta_{\bf r}\ ,
\end{displaymath}
Fourier transformation of the discrete variables $m_{\bf r,r'}$,

\begin{eqnarray}\nonumber
m_{{\bf r,r+a}_\alpha} &=& \frac{1}{\sqrt{N}}\sum_{\bf q}\
e^{-i{\bf q(r+a}_\alpha/2)} m^{\alpha}_{\bf q}\ ,\qquad \alpha =
x, y
\\\nonumber m^{\alpha}_{\bf q} &=& \frac{1}{\sqrt{N}}\sum_{\bf r}\ e^{i{\bf
q(r+a}_\alpha/2)} m_{{\bf r,r+a}_\alpha}\ ,\qquad \alpha = x, y\ ,
\end{eqnarray}
and the shift of the Fourier transform of the spin variable:

\begin{displaymath}
\theta_{\bf k} = \varphi_{\bf k} - 2\pi i
\frac{\sum_{\alpha}K_{\alpha}({\bf k})m^{\alpha}_{\bf
k}}{\sum_{\gamma}K^2_{\gamma}({\bf k})}
\end{displaymath}
with $K_{\alpha}({\bf k})\equiv 2\sin{\frac{k_{\alpha}a}{2}}$.

Applying this scheme to the diluted Hamiltonian (\ref{Hamilt_Villain}) we find
that again, as in the pure case, the two parts - the spin-wave 
and the vortex one - can be distinguished
but now a third term appears which 
depends both on the spin and vortex degrees of freedom. 
Thus the Hamiltonian is made of three terms as:

\begin{equation}\label{three_parts}
H_{\mathrm{Vill}}^{\mathrm{dil}} =
H_{\mathrm{SW}}^{\mathrm{dil}} + H_{\mathrm{Vort}}^{\mathrm{dil}} + H_{\mathrm{SW,Vort}}^{\mathrm{dil}}\ .
\end{equation}
Of course, the cross-term, $H_{\mathrm{SW,Vort}}^{\mathrm{dil}}$, vanishes in the pure model limit: $c_{\bf r}=1, {\bf r}=1,...,N$. It has the form:

\begin{eqnarray}\nonumber
H_{\mathrm{SW,Vort}}^{\mathrm{dil}} &=& 4\pi J\sum_{\bf k,k'}\rho({\bf
k+k'})\bigg(L_x({\bf k+k'})K_x({\bf k})K_y({\bf
k'})\\\label{SW_Vort}
&-& L_y({\bf k+k'})K_y({\bf k})K_x({\bf
k'})\bigg)\left(K^2_x({\bf k'})+K^2_y({\bf k'})
\right)^{-1}\varphi_{\bf k}q_{\bf
k'}\ ,
\end{eqnarray}
where $q_{\bf k}=i(K_y({\bf k})m^x_{\bf k}-K_x({\bf k})m^y_{\bf k})$ 
are the Fourier transforms of the vortex charges, 
$L_\alpha({\bf k})=\cos\frac{k_\alpha a}{2}$, $K_{\alpha}({\bf k})$
 
was defind before, and

\begin{equation}\label{rho}
\rho({\bf q}) = \frac{1}{N}\sum_{\bf r}e^{i{\bf qr}}(1-c_{\bf r})\ .
\end{equation}
The parameter (\ref{rho}) characterizes the strength of
dilution~\cite{Berche03}. When there are
no impurities in the system one gets $\rho({\bf k+k'})=0$ and 
thus $H_{\mathrm{SW,Vort}}^{\mathrm{dil}}=0$. The spin-wave and 
vortex parts of the Hamiltonian also contain terms which depend
on $\rho$ and
which turn to zero in the pure case as well.

The spin-wave term:

\begin{eqnarray}\nonumber
H_{\mathrm{SW}}^{\mathrm{dil}} &=& \frac{J}{2}\sum_{\bf
k}\sum_{\alpha}K^2_{\alpha}({\bf k})\varphi_{\bf k}\varphi_{\bf
-k}\\\label{SW_part} &+& 
J\sum_{\bf k,k'}\rho({\bf
k+k'})\left(\sum_{\alpha}L_\alpha({\bf k+k'})K_{\alpha}({\bf
k})K_{\alpha}({\bf k'})\right)\varphi_{\bf k}\varphi_{\bf
k'}\ ,
\end{eqnarray}
contains the ``pure" spin-wave Hamiltonian (first term) and a contribution of the dilution that vanishes when $\rho({\bf k+k'})=0$. The dilution contribution naturally has exactly the same form as was found in~\cite{Berche03}.

The new result here is the form of 
the vortex energy in the presence of non-magnetic dilution:

\begin{eqnarray}\label{Vort_part}
H_{\mathrm{Vort}}^{\mathrm{dil}} &=&
2\pi J\sum_{\bf k\neq 0}\frac{q_{\bf k}q_{\bf
-k}}{\sum_{\gamma}K^2_{\gamma}({\bf k})}
+
4\pi J\sum_{\bf k,k'}\rho({\bf
k+k'})
\\\nonumber
&\times&\left(\frac{L_x({\bf k+k'})K_y({\bf k})K_y({\bf
k'}) + L_y({\bf k+k'})K_x({\bf k})K_x({\bf
k'})}{(K^2_x({\bf k})+K^2_y({\bf k}))(K^2_x({\bf k'})+K^2_y({\bf
k'}))}\right)q_{\bf k}q_{\bf k'}\ .
\end{eqnarray}
Again one has the vortex interactions similar to those of 
the pure Villain model (first term) 
while the dilution effect is represented by the second term.
It is known for the ``pure" Villain term that:

\begin{eqnarray}\nonumber
H_{\mathrm{Vort}}^{\mathrm{pure}} &=&
2\pi^2 J\sum_{\bf k\neq 0}\frac{q_{\bf k}q_{\bf
-k}}{\sum_{\gamma}K^2_{\gamma}({\bf k})} 
\\
&\simeq& -2\pi\sum_{\bf R,R'}q({\bf R})q({\bf R'})\ln(|{\bf R-R'}|/a) + \pi^2 J\sum_{\bf R}\left(q({\bf R})\right)^2\ ,
\end{eqnarray}
where ${\bf R}$, ${\bf R'}$ span the sites of the dual lattice and $q({\bf R})$-s are the vortex charges or winding numbers~\cite{Villain75}.

Note, that in order to present the dilution contributions of the Hamiltonian (\ref{three_parts})
in an easily readable form we 
have not included into Eq.(\ref{SW_Vort}), (\ref{SW_part}) and (\ref{Vort_part}) the terms quadratic in $\rho$. Anyway, one can neglect them in the approximation of a weak dilution (see~\cite{Berche03}) which is the case here.

The expression presented in (\ref{SW_Vort})
and especially the form of
the impurity contribution to the vortex part of the Hamiltonian,
Eq.(\ref{Vort_part}), can further serve to analyze the impact of
dilution on the peculiarities of the BKT transition. Let us first
find an approximation that would correspond to the ``smeared"
impurity density approximation of Section 3.

Imagine that the fraction $1-c$ of sites is removed in such a way that the vacancies form some regular structure, then 

\begin{equation}\label{rho_mean}
\rho({\bf k+k'})=(1-c)\delta_{{\bf k+k'},0}\ .
\end{equation}
Of course, with random dilution this is not the case,
but, as was argued in~\cite{Kapikranian07}, (\ref{rho_mean}) can be 
considered as the zero approximation when one neglects the random 
fluctuations (homogeneous dilution). Moreover, (\ref{rho_mean}) is 
 the disorder averaged value of $\rho$. We will see that 
this replacement with its average value corresponds to the ``smeared" 
impurity density approximation of Section 3. In this case we obtain 
the vortex energy:

\begin{equation}\label{Vort_Enrg_Apprxm}
H_{\mathrm{Vort}}^{\mathrm{dil}} =
2\pi \left[1-2(1-c)\right] J\sum_{\bf k\neq 0}\frac{q_{\bf k}q_{\bf
-k}}{\sum_{\gamma}K^2_{\gamma}({\bf k})}\ .
\end{equation}
As a consequence the energy of a neutral vortex pair is $E^{\mathrm{dil}}_{\mathrm{int}}(R) = [1-2(1-c)] 2\pi J\ln(R/a)$, exactly the same result 
as what was found from the topological defect theory 
approach under the assumption of the ``smeared" density of 
vacancies. This leads of course to the same estimate of 
the critical temperature as well.

\section{Conclusions}\label{Concl}

We have exploited two different approaches to account for
the influence of 
quenched impurities on the vortices in the two-dimensional $XY$ model: 
in the frame of the topological defects phenomenological theory and from
the Villain model Hamiltonian. The interaction of a vacancy with vortices 
in the theory of topological defects was found to be attractive, in good 
accordance with other works on this subject~\cite{PereiraEtAl03,Wysin03}.
In order to estimate the critical temperature change we used an approach 
based on the vortex-pair dissociation mechanism of the BKT transition. 
The ``smeared" impurity density approximation leads to the same
predictions for the critical temperature within
the two approaches 
(topological defects theory and the Villain model). The dependence
of the  transition temperature on the magnetic sites concentration,
$T_{\mathrm{BKT}}^{\mathrm{dil}}(c)$, obtained under the assumption of a 
weak dilution, however gives a percolation threshold which differs 
from the known real site
percolation threshold for a $2D$ square lattice. 
Comparing with the currently available Monte Carlo results~\cite{WysinEtAl05} 
which unfortunately are not accurate
enough to make reliable conclusion about the 
weak dilution range, we recover the linear
character of the critical 
temperature decrease close to $c=1$.

%%%%%%%%%%%%%%%%%%%%%%%%%%%%

\section*{Acknowledgements}

We acknowledge the CNRS-NAS Franco-Ukrainian bilateral exchange program and
Thierry Platini for the interesting discussions. Yu. H. acknowledges support of the Austrian FWF project 19583-PHY.

\end{document}